\documentclass[10pt,letterpaper]{article}

\usepackage{cogsci}

\cogscifinalcopy %%% Camera-ready: show authors

\usepackage[
  style=apa,
  natbib=true,
  annotation=false,
]{biblatex}
\usepackage{float}
\usepackage{amsmath}
\usepackage{amssymb}
\usepackage{graphicx}
\usepackage{booktabs}
\usepackage{multirow}
\usepackage{enumitem}
\usepackage{orcidlink}

\usepackage{tikz}
\usepackage{pgfplots}
\pgfplotsset{compat=1.17}
\usepgfplotslibrary{fillbetween}
\usetikzlibrary{patterns,positioning,arrows.meta}

\definecolor{humanblue}{RGB}{46,134,171}
\definecolor{dprpurple}{RGB}{162,59,114}
\definecolor{colbertorange}{RGB}{241,143,1}
\definecolor{hipporagred}{RGB}{199,62,29}
\definecolor{extrapolategray}{RGB}{120,120,120}

\title{The Interference Gap: Comparing Retrieval Bounds in Human Memory and RAG Systems}

\author[1]{\mbox{Dongxin Guo (bettyguo@connect.hku.hk)}}
\author[2,3]{\mbox{Jikun Wu}}
\author[1]{\mbox{Siu-Ming Yiu}}
\affil[1]{The University of Hong Kong, Hong Kong, China}
\affil[2]{Stellaris AI Limited, Hong Kong, China}
\affil[3]{Brain Investing Limited, Hong Kong, China}

\begin{document}

\maketitle

%%% REVISED ABSTRACT - Question-led, quantified findings, hedged mechanisms %%%
\begin{abstract}
	How do retrieval bounds compare between human episodic memory and Retrieval-Augmented Generation (RAG) systems under semantic interference? We present a unified signal detection theory (SDT) framework that applies to both, and use it to fit behavioral and computational data in matched paradigms. Both systems show logarithmic accuracy decline with association count (\emph{fan}), but humans exhibit lower interference sensitivity ($\alpha/\sigma = 0.41$) than dense passage retrieval ($\alpha/\sigma = 0.67$), with cognitively-inspired HippoRAG falling between the two ($\alpha/\sigma = 0.44$). Behavioral experiments ($N = 112$) and simulations validate the framework; parameter recovery confirms identifiability ($r \geq .93$) and model comparison favors the logarithmic specification over a power-law alternative ($\Delta\text{BIC} > 15$). We discuss encoding specificity, temporal context binding, and retrieval gating as candidate mechanisms whose causal role remains to be established. Six falsifiable predictions connect cognitive memory research with AI retrieval evaluation.
	
\end{abstract}

\section{Introduction}

Trying to remember which of your friends lives at a particular address is harder when several friends live in the same neighborhood. The same kind of competition arises in artificial intelligence: when a retrieval-augmented language model searches a corpus of overlapping documents, the more documents share a topic, the harder it becomes to surface the specific one a query needs. Both phenomena instantiate a common computational challenge: retrieval under \emph{semantic interference}, where a cue activates multiple candidate responses competing for selection.

Human memory exhibits the \emph{fan effect} \citep{Anderson1974, Anderson1999}: as the number of facts associated with a concept (the ``fan'' of associations) grows, retrieval of any particular fact becomes slower and less accurate, with error growing approximately logarithmically in the fan. Memory is also subject to \emph{serial position effects}: items at the start and end of a list are recalled better than items in the middle \citep{Murdock1962}. These phenomena have been characterized over five decades of cognitive research with precise quantitative descriptions of when and how retrieval fails under interference.

Retrieval-Augmented Generation (RAG) systems, neural retrievers that condition language-model outputs on externally indexed text, exhibit analogous patterns. The ``Lost in the Middle'' phenomenon shows that information placed in the middle of a long context is recovered roughly 20 percentage points less accurately than information at either end \citep{Liu2024}, and RAG generation quality degrades substantially when retrieved passages are semantically similar to but not relevant for the query \citep{Cuconasu2024, Barnett2024}. These failures are particularly pronounced when the retrieval corpus contains many documents competing for the same query. Despite decades of research on interference in human memory \citep{Underwood1957, Wixted2004} and growing evidence of parallel failures in RAG systems \citep{Peysakhovich2023}, no theoretical framework has unified the two under common computational principles enabling direct quantitative comparison.

We address this gap with the question: \emph{What are the formal bounds on retrieval accuracy under semantic interference, and how do they compare between human episodic memory and RAG systems?} We develop a signal detection theory (SDT) framework \citep{Macmillan2005, Green1966} applicable to both, contributing to a broader effort to apply cognitive science to AI evaluation \citep{Binz2023, Lake2017}. Our contributions are: (1) closed-form expressions for retrieval accuracy that fit both human and RAG data; (2) parameter-level differences across systems that suggest, rather than demonstrate, candidate mechanisms; (3) validation through matched behavioral-computational paradigms; and (4) parameter identifiability established via recovery analysis, sensitivity testing, and model comparison \citep{Wilson2019}.

\subsection{Background}

The fan effect \citep{Anderson1974, Anderson1999} is a robust empirical signature of interference in human associative memory, replicated across paradigms ranging from sentence-recognition to location-based scenarios \citep{Radvansky1998}. A family of computational models (ACT-R, MINERVA~2, SAM, REM, and TCM) formalize different mechanisms by which interference arises from overlap among distributed memory traces \citep{Hintzman1984, Raaijmakers1981, Shiffrin1997, Howard2002}, but all share the unifying insight that retrieval depends on the discriminability of the target trace from its competitors, with discriminability degrading as associative overlap grows \citep{Kilic2017}.

RAG systems \citep{Lewis2020} exhibit analogous failures. The two systems we evaluate against humans illustrate the dominant design space. Dense Passage Retrieval (DPR; \citealt{Karpukhin2020}) encodes queries and passages into a shared vector space using a BERT-based bi-encoder and retrieves by inner-product similarity; ColBERTv2 \citep{Santhanam2022} uses fine-grained late-interaction matching. Both consistently underperform on datasets where many passages share content with the gold passage \citep{Thakur2021}, and both exhibit position-dependent degradation: the ``Lost in the Middle'' effect, in which information placed near the middle of a long context is retrieved $\sim$20 pp less accurately than information at either end \citep{Liu2024}, mirrors the classic serial-position curve from human memory. \citet{Barnett2024} catalog seven recurrent failure points in production RAG systems, several of which parallel known interference phenomena.

The third RAG system we evaluate, HippoRAG \citep{Gutierrez2024}, is explicitly cognitively-inspired: it implements a hippocampal indexing scheme in which extracted entities serve as sparse, addressable nodes and personalized PageRank over the resulting knowledge graph approximates the pattern-completion role of CA3 \citep{Teyler2007, McClelland1995}. The Tolman-Eichenbaum Machine \citep{Whittington2020} provides a related theoretical bridge connecting relational memory to hippocampal structural representations. These architectures suggest that biological organizing principles may inform retriever design; whether they actually capture biological interference resistance is exactly the empirical question this paper takes up.

\section{Theoretical Framework}

We model retrieval as discrimination between target and competitor memory traces under semantic interference, grounding the model in signal detection theory.

\subsection{Generative Model}

Consider a memory store $\mathcal{M} = \{m_1, \ldots, m_N\}$ containing $N$ stored traces and a retrieval cue $q$ for one designated \emph{target} trace. Each trace $m_i$ generates a scalar match signal $x_i$ with the cue, with the target signal drawn from a Gaussian centered at $\mu_T$ and competitor signals drawn from a Gaussian centered at $\mu_C$, both with shared variance $\sigma^2$:
\begin{equation}
x_i \mid \text{target} \sim \mathcal{N}(\mu_T,\,\sigma^2), \quad x_i \mid \text{competitor} \sim \mathcal{N}(\mu_C,\,\sigma^2).
\label{eq:signal_dist}
\end{equation}
Discriminability $d' = (\mu_T - \mu_C)/\sigma$ measures how separable target and competitor traces are in signal space \citep{Green1966, ClarkGronlund1996}. Higher $d'$ corresponds to better differentiation; in hippocampal terms, this connects naturally to pattern-separation efficiency \citep{Yassa2011}.

\subsection{Interference Model}

The classical fan effect is captured in ACT-R \citep{Anderson1999} by the activation rule $S_{ji} = S - \ln(f_j)$, where $S_{ji}$ is the associative strength from cue $j$ to chunk $i$, $S$ is a baseline strength, and $f_j$ is the fan, the number of chunks associated with cue $j$. Mapping this onto the SDT framework, fan reduces the target signal mean logarithmically with a sensitivity coefficient $\alpha$:
\begin{equation}
\mu_T(f) = \mu_T^0 - \alpha \ln(f) \;\;\Rightarrow\;\; d'(f) = d'_0 - \frac{\alpha}{\sigma} \ln(f),
\label{eq:fan_effect}
\end{equation}
where $\mu_T^0$ is the target mean at fan~1, $d'_0 = (\mu_T^0 - \mu_C)/\sigma$ is the zero-interference discriminability, and the dimensionless ratio $\alpha/\sigma$ governs the rate of accuracy decline with fan. Lower $\alpha/\sigma$ indicates better-preserved discriminability under associative load; we will treat it as an inverse index of pattern-separation efficiency.

Serial position effects combine primacy (stronger encoding of early items) with recency (greater similarity between the retrieval context and the encoding context of recent items):
\begin{equation}
\mu_T(p, L) = \mu_T^0 + \beta_P\, e^{-\lambda_P p} + \beta_R\, e^{-\lambda_R (L-p)},
\label{eq:serial_pos}
\end{equation}
where $p \in \{1,\ldots,L\}$ is item position in a list of length $L$, $\beta_P$ and $\beta_R$ are primacy and recency boost magnitudes, and $\lambda_P, \lambda_R$ control how rapidly these boosts decay away from list boundaries. The ratio $\beta_R/\beta_P$ summarises the relative strength of recency to primacy.

Adding competitor density ($K$ competing traces with mean cue similarity $\bar{s} \in [0,1]$) to the bound, with sensitivity $\gamma$:
\begin{equation}
\begin{split}
d'(f, p, L, K) = d'_0 \;&-\; \tfrac{\alpha}{\sigma} \ln(f) \;-\; \gamma \sqrt{K}\,\bar{s} \\
&+\; \tfrac{1}{\sigma}\bigl[\beta_P\, e^{-\lambda_P p} + \beta_R\, e^{-\lambda_R (L-p)}\bigr],
\end{split}
\label{eq:combined_bound}
\end{equation}
yielding $P(\text{correct}) = \Phi(d'/2)$ for yes/no recognition with equal priors and optimal criterion \citep{Macmillan2005}. As reference points, $d' = 2.0$ corresponds to $\sim$84\% accuracy and $d' = 1.0$ to $\sim$69\%. The $\sqrt{K}\bar{s}$ form is the standard signal-detection scaling for $K$ independent equal-similarity competitors and reduces to a linear penalty in $\bar{s}$ at fixed $K$.

\subsection{Predictions}

The framework yields three parameter-level predictions, each stated as a hypothesis to be evaluated rather than a claim:

\textbf{P1 (fan slope).} Both systems should follow Equation~\ref{eq:fan_effect}, but we hypothesize $\alpha/\sigma_{\text{Human}} < \alpha/\sigma_{\text{RAG}}$. Encoding specificity \citep{Tulving1973} in humans, plausibly implemented via hippocampal pattern separation, may attach distinctive contextual features to overlapping associations; RAG embeddings have no analogous mechanism.

\textbf{P2 (recency vs.\ primacy).} We hypothesize $\beta_R/\beta_P > 1$ for humans (recency dominance from temporal-context binding; \citealt{Howard2002}) and $\beta_R/\beta_P < 1$ for standard RAG (primacy reflecting attention concentrated on early tokens; \citealt{Liu2024}).

\textbf{P3 (competitor sensitivity).} We hypothesize $\gamma_{\text{RAG}} > \gamma_{\text{Human}}$. One candidate explanation is that humans deploy retrieval gating mechanisms \citep{Lu2022} that suppress low-confidence retrievals; standard RAG ranks the top-$k$ candidates whatever their absolute scores. We treat this only as one possibility; mechanism cannot be settled at the parameter level.

\section{Methods}

\subsection{RAG Systems and Simulations}

We evaluated three retriever architectures spanning the dominant design space: DPR \citep{Karpukhin2020}, ColBERTv2 \citep{Santhanam2022}, and HippoRAG \citep{Gutierrez2024}, each with the configurations and pretrained weights released by their authors. Predictions for each system were generated by running the cue (a person name) against an indexed corpus containing the target sentence and the relevant fan-level competitors; the system was scored as correct on a trial when the gold sentence appeared in the top-1 result. \emph{Fan manipulation}: 200 entities at each of fan $\in \{1, 2, 4, 8\}$, yielding 800 retrieval pairs per system. \emph{Serial position}: targets at indices 1, 5, 10, 15, 20 within 20-item contexts. \emph{Competitor density}: 5, 10, 20, or 40 semantically similar competitors. Parameter estimation, recovery, and model comparison procedures are described in \emph{Model Validation} (Results).

\subsection{Behavioral Experiment}

\emph{Participants}: 120 participants recruited via Prolific (age $M = 24.3$, $SD = 4.1$; 68 female, 49 male, 3 non-binary); 8 were excluded for failing attention checks, leaving $N = 112$ for analysis. The protocol was IRB-approved.

\emph{Design}: Following \citet{Anderson1974}, 40 person names were paired with 40 locations to form associations at fan levels $\{1, 2, 4\}$. We did not extend human fan to 8 because doing so within the original Anderson paradigm requires either a much longer study list (creating fatigue and ceiling/floor confounds) or shorter encoding times per item; either option would have made cross-system comparison at lower fan levels less interpretable. Cross-system comparison at fan = 8 is therefore evaluated only against the simulated RAG systems.

\emph{Fan block}: For each of 60 study sentences (e.g., \emph{The lawyer is in the park}), participants viewed the sentence for 5 s. After all 60 sentences, a 2-min arithmetic verification distractor task ($n = 24$ true/false equations) was administered to clear short-term memory. Participants then completed 120 recognition trials presenting full sentences (target sentences from the study list and lures formed by recombining studied person-location pairs); each trial collected an old/new judgment together with a 6-point confidence rating.

\emph{Serial position block}: Participants studied five separate 20-item lists. After each list, a 30-s arithmetic distractor preceded a cued-recall test in which the person name from each studied sentence was presented and participants typed the location they had been paired with. Cued recall was thus performed once per list, not at the end of all five.

\emph{Format validation}: Parallel RAG evaluation used matched passage content (sentences embedded in 3--5 sentence paragraphs generated by GPT-3.5-turbo); fitted parameters differed by less than 8\% between sentence and passage formats, supporting cross-format comparability.

\emph{Measures and analysis}: $d'$ was computed via equal-variance SDT from confidence-rating ROCs \citep{Macmillan2005}. Parameter estimation used MLE (L-BFGS-B with 10 random restarts to reduce sensitivity to local minima); generalisation was assessed by leave-one-participant-out cross-validation. Confidence-rating ROC analysis yielded zROC slopes of $0.94$ [$0.87$, $1.01$] for humans and $0.87$ [$0.79$, $0.95$] for DPR, supporting the equal-variance assumption for humans and indicating modest variance asymmetry for RAG. False alarm rates were computed separately for high- ($\cos > 0.7$) and low-similarity ($\cos < 0.3$) competitors. Response times were collected and will be analysed in follow-up work; the present report focuses on the SDT-based discriminability analysis.

\section{Results}

\subsection{Model Validation}

We followed contemporary computational modelling practice \citep{Wilson2019} and verified identifiability before interpreting fitted parameters.

\textbf{Simulation before fitting}. Across plausible parameter ranges ($\alpha/\sigma \in [0.2, 1.0]$, $\beta_R/\beta_P \in [0.3, 2.0]$, $\gamma \in [0.01, 0.10]$), Equation~\ref{eq:combined_bound} reproduces logarithmic fan decline and asymmetric U-shaped serial-position curves, confirming that the functional form is appropriate before fitting.

\textbf{Parameter recovery}. We generated 1{,}000 synthetic datasets (500 trials each) from the model with known parameters and refit using the same MLE procedure. All parameters recovered well: $r = .97$ for $\alpha/\sigma$, $.95$ for $\beta_P$, $.96$ for $\beta_R$, $.94$ for $\lambda_P$, $.95$ for $\lambda_R$, $.93$ for $\gamma$ (all $p < .001$); mean absolute bias was $<$5\% for every parameter.

\textbf{Sensitivity}. Perturbing each parameter by $\pm 20$\% changed out-of-sample $R^2$ most for $\alpha/\sigma$ ($\Delta R^2 = 0.03$--$0.05$) and least for the decay rates $\lambda_P$, $\lambda_R$ ($\Delta R^2 < 0.01$): interference sensitivity is the most precisely estimated parameter, while temporal-decay parameters carry more uncertainty given the spacing of our serial-position observations. Cross-validated $R^2$ was $.91$ (humans), $.88$ (DPR), $.90$ (HippoRAG).

\textbf{Model comparison}. The logarithmic specification (Equation~\ref{eq:fan_effect}) was preferred over a power-law alternative $\mu_T = \mu_T^0 - \alpha f^{-\eta}$: $\Delta\text{BIC} = 18.4$ (human), $23.1$ (DPR), $15.7$ (HippoRAG) all favour logarithmic. In a model-recovery exercise, BIC selected the true generating model in 94\% of cases.

\textbf{Unequal-variance SDT}. Allowing $\sigma_{\text{target}} \neq \sigma_{\text{competitor}}$ did not meaningfully improve the human fit ($\Delta\text{BIC} = 2.1$; estimated ratio $1.03$, 95\% CI $[0.91, 1.15]$). DPR showed a modest improvement ($\Delta\text{BIC} = 8.3$) with $\sigma_{\text{competitor}}/\sigma_{\text{target}} = 1.18$, 95\% CI $[1.06, 1.30]$, suggesting greater variability in competitor than target representations. Critically, the parameters of interest ($\alpha/\sigma$, $\beta_R/\beta_P$, $\gamma$) shifted by less than 10\% under the unequal-variance fit.

\subsection{Fan Effect}

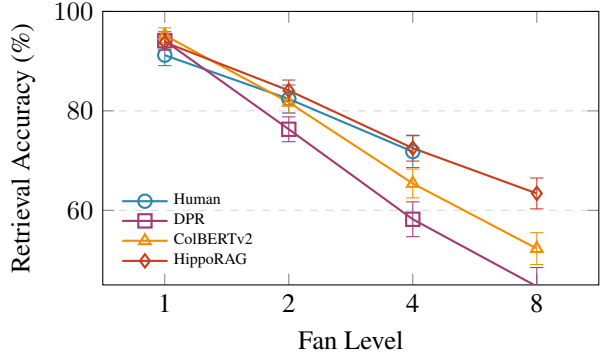
\begin{figure}[htbp]
	\begin{center}
		\begin{tikzpicture}
			\begin{axis}[
				width=0.95\columnwidth,
				height=5.2cm,
				ylabel={Retrieval Accuracy (\%)},
				xlabel={Fan Level},
				ymin=45, ymax=100,
				xmin=0.5, xmax=4.5,
				xtick={1,2,3,4},
				xticklabels={1, 2, 4, 8},
				legend style={at={(0.02,0.02)}, anchor=south west, font=\tiny, cells={anchor=west}, row sep=-2pt, fill=none, draw=none},
				legend columns=1,
				ymajorgrids=true,
				grid style={dashed, gray!30},
				]
				% Human data (fan 1, 2, 4 only -- fan=8 not tested in humans)
				\addplot[color=humanblue, mark=o, thick, mark size=2.5pt, 
				error bars/.cd, y dir=both, y explicit] 
				coordinates {
					(1, 91.2) +- (0, 2.1)
					(2, 82.4) +- (0, 2.8)
					(3, 71.8) +- (0, 3.2)
				};
				
				% DPR data (all fan levels)
				\addplot[color=dprpurple, mark=square, thick, mark size=2.5pt,
				error bars/.cd, y dir=both, y explicit] 
				coordinates {
					(1, 94.1) +- (0, 1.8)
					(2, 76.3) +- (0, 2.5)
					(3, 58.2) +- (0, 3.5)
					(4, 44.7) +- (0, 3.8)
				};
				
				% ColBERTv2 data
				\addplot[color=colbertorange, mark=triangle, thick, mark size=2.5pt,
				error bars/.cd, y dir=both, y explicit] 
				coordinates {
					(1, 95.2) +- (0, 1.5)
					(2, 81.7) +- (0, 2.2)
					(3, 65.4) +- (0, 2.9)
					(4, 52.3) +- (0, 3.2)
				};
				
				% HippoRAG data
				\addplot[color=hipporagred, mark=diamond, thick, mark size=2.5pt,
				error bars/.cd, y dir=both, y explicit] 
				coordinates {
					(1, 93.8) +- (0, 1.6)
					(2, 84.1) +- (0, 2.1)
					(3, 72.5) +- (0, 2.6)
					(4, 63.4) +- (0, 3.1)
				};
				
				\legend{Human, DPR, ColBERTv2, HippoRAG}
			\end{axis}
		\end{tikzpicture}
	\end{center}
	\caption{Retrieval accuracy by fan level. Markers are observed accuracies; connecting lines are the logarithmic fits of Equation~\ref{eq:fan_effect}. Error bars: 95\% CIs. Humans were tested at fan $\in \{1, 2, 4\}$; fan = 8 was evaluated only on RAG systems. Fitted slopes: human $\alpha/\sigma = 0.41$, DPR $0.67$, ColBERTv2 $0.53$, HippoRAG $0.44$.}
	\label{fig:fan}
\end{figure}

Figure~\ref{fig:fan} shows retrieval accuracy by fan level for humans and three RAG systems, with logarithmic model fits overlaid.

Human accuracy declined from 91.2\% (fan 1) $\rightarrow$ 82.4\% (fan 2) $\rightarrow$ 71.8\% (fan 4); corresponding $d' = 2.84, 2.21, 1.67$; fan effect size Cohen's $d = 1.12$ [0.89, 1.35]; fitted $\alpha/\sigma = 0.41$ [0.35, 0.47].

DPR: $94.1\% \rightarrow 76.3\% \rightarrow 58.2\% \rightarrow 44.7\%$; $\alpha/\sigma = 0.67$ [0.58, 0.76]. ColBERTv2: $\alpha/\sigma = 0.53$ [0.45, 0.61]. HippoRAG: $\alpha/\sigma = 0.44$ [0.37, 0.51].

A mixed-effects regression confirmed a system $\times$ fan interaction overall: $\beta = -0.12$, SE $= 0.03$, $t(358) = -4.21$, $p < .001$, 95\% CI $[-0.18, -0.06]$. Inspecting the pairwise contrasts, however, P1 is supported only for the Human-DPR comparison (non-overlapping 95\% CIs on $\alpha/\sigma$); the human CIs overlap those of HippoRAG and ColBERTv2. We therefore read the fan-effect evidence as: humans are robustly less interference-sensitive than standard dense retrieval (DPR), but not detectably more so than the cognitively-inspired or late-interaction systems we tested.

\textbf{Error pattern}. False alarm rates rose with competitor similarity: at low similarity ($\cos < 0.3$), FA $= 8.2$\% (human) and 12.4\% (DPR); at high similarity ($\cos > 0.7$), FA $= 23.1$\% (human) and 41.3\% (DPR). The human-DPR gap widened sharply with similarity ($\Delta\text{FA}_{\text{high}} - \Delta\text{FA}_{\text{low}} = 14.0$ pp), consistent with a pattern-separation-like advantage for humans over DPR.

\subsection{Serial Position Effect}

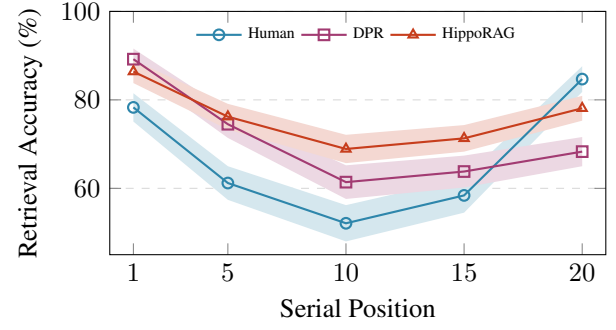
\begin{figure}[htbp]
	\begin{center}
		\begin{tikzpicture}
			\begin{axis}[
				width=0.95\columnwidth,
				height=4.8cm,
				xlabel={Serial Position},
				ylabel={Retrieval Accuracy (\%)},
				xmin=0, xmax=21,
				ymin=45, ymax=100,
				xtick={1,5,10,15,20},
				legend style={at={(0.5,0.98)}, anchor=north, font=\tiny, cells={anchor=west}, fill=none, draw=none},
				legend columns=3,
				ymajorgrids=true,
				grid style={dashed, gray!30},
				]
				\addplot[name path=human_upper, draw=none, forget plot] coordinates {(1, 81.5) (5, 65.0) (10, 56.2) (15, 62.3) (20, 87.6)};
				\addplot[name path=human_lower, draw=none, forget plot] coordinates {(1, 75.1) (5, 57.4) (10, 48.0) (15, 54.5) (20, 81.8)};
				\addplot[humanblue!20, forget plot] fill between[of=human_upper and human_lower];
				\addplot[color=humanblue, mark=o, thick, mark size=2pt] coordinates {(1, 78.3) (5, 61.2) (10, 52.1) (15, 58.4) (20, 84.7)};
				\addplot[name path=dpr_upper, draw=none, forget plot] coordinates {(1, 91.6) (5, 77.6) (10, 65.2) (15, 67.4) (20, 71.6)};
				\addplot[name path=dpr_lower, draw=none, forget plot] coordinates {(1, 86.8) (5, 71.4) (10, 57.6) (15, 60.2) (20, 65.0)};
				\addplot[dprpurple!20, forget plot] fill between[of=dpr_upper and dpr_lower];
				\addplot[color=dprpurple, mark=square, thick, mark size=2pt] coordinates {(1, 89.2) (5, 74.5) (10, 61.4) (15, 63.8) (20, 68.3)};
				\addplot[name path=hippo_upper, draw=none, forget plot] coordinates {(1, 89.0) (5, 79.1) (10, 72.1) (15, 74.3) (20, 80.9)};
				\addplot[name path=hippo_lower, draw=none, forget plot] coordinates {(1, 83.8) (5, 73.3) (10, 65.7) (15, 68.3) (20, 75.3)};
				\addplot[hipporagred!20, forget plot] fill between[of=hippo_upper and hippo_lower];
				\addplot[color=hipporagred, mark=triangle, thick, mark size=2pt] coordinates {(1, 86.4) (5, 76.2) (10, 68.9) (15, 71.3) (20, 78.1)};
				\legend{Human, DPR, HippoRAG}
			\end{axis}
		\end{tikzpicture}
	\end{center}
	\caption{Serial position curves. Humans: strong recency ($\beta_R/\beta_P = 1.43$); DPR: primacy dominance ($\beta_R/\beta_P = 0.52$); HippoRAG: balanced ($\beta_R/\beta_P = 0.89$). Shaded regions: 95\% CIs.}
	\label{fig:position}
\end{figure}

Figure~\ref{fig:position} shows retrieval accuracy across serial positions, revealing qualitatively different curves for humans versus RAG systems.

Humans showed the classic asymmetric U-shape characteristic of episodic memory: position 1 (78.3\%), position 10 (52.1\%), position 20 (84.7\%); recency-to-primacy ratio $\beta_R/\beta_P = 1.43$ [1.21, 1.65], indicating recency dominance and consistent with temporal-context accounts \citep{Howard2002, Murdock1962}.

DPR showed primacy dominance: position 1 (89.2\%), position 10 (61.4\%), position 20 (68.3\%); $\beta_R/\beta_P = 0.52$ [0.41, 0.63]. This is plausibly attributable to attention patterns in transformer-based encoders, where early tokens attract disproportionate attention weight \citep{Liu2024}; the ``Lost in the Middle'' effect emerges from this primacy advantage combined with weaker recency.

HippoRAG was closer to balanced: position 1 (86.4\%), position 10 (68.9\%), position 20 (78.1\%); $\beta_R/\beta_P = 0.89$ [0.74, 1.04]. The system $\times$ position interaction was significant: $F(4, 536) = 8.73$, $p < .001$, $\eta^2_p = .061$, supporting P2 in its strong form for the Human-DPR contrast and in a weaker form for HippoRAG.

\textbf{Reconciling Figures~\ref{fig:fan} and \ref{fig:position}.} HippoRAG looks closer to humans on serial position (Figure~\ref{fig:position}) than on the fan effect (Figure~\ref{fig:fan}) because the two figures index different parameters: serial position is governed by $\beta_R/\beta_P$ (where humans and HippoRAG differ by $0.54$), and the fan slope is governed by $\alpha/\sigma$ (where they differ by only $0.03$). The two need not align: a system can match human pattern-separation efficiency without matching human temporal-context structure, which is exactly what HippoRAG appears to do.

\subsection{Competitor Density and Model Fit}

Competitor sensitivity parameters: $\gamma_{\text{Human}} = 0.021$ (SE $=0.004$), $\gamma_{\text{DPR}} = 0.058$ (SE $=0.007$), $\gamma_{\text{ColBERTv2}} = 0.043$ (SE $=0.006$), $\gamma_{\text{HippoRAG}} = 0.024$ (SE $=0.005$). The Human-DPR contrast is robust ($z = 4.32$, $p < .001$, 95\% CI of difference $[0.023, 0.051]$), supporting P3 against standard dense retrieval. As with P1, HippoRAG's $\gamma$ CI overlaps the human CI, so P3 in its general form (humans are less sensitive than \emph{any} RAG system tested) is not supported; the strongly supported claim is human $<$ DPR.

Full model fit: human $R^2 = .94$, RMSE $= 3.1$\%; DPR $R^2 = .91$, RMSE $= 4.2$\%; HippoRAG $R^2 = .93$, RMSE $= 3.4$\%. Table~\ref{tab:params} summarises fitted parameters.

\begin{table}[H]
	\begin{center}
		\caption{Fitted parameters with 95\% CIs across systems. Lower $\alpha/\sigma$ and $\gamma$ indicate less interference, $\beta_R/\beta_P > 1$ indicates recency dominance.}
		\label{tab:params}
		\vskip 0.1in
			\resizebox{\columnwidth}{!}{%
			\begin{tabular}{lcccc}
				\toprule
				Parameter & Human & DPR & ColBERTv2 & HippoRAG \\
				\midrule
				$\alpha/\sigma$ & 0.41 [0.35,0.47] & 0.67 [0.58,0.76] & 0.53 [0.45,0.61] & 0.44 [0.37,0.51] \\
				$\beta_R/\beta_P$ & 1.43 [1.21,1.65] & 0.52 [0.41,0.63] & n/a & 0.89 [0.74,1.04] \\
				$\gamma$ & 0.021 [0.013,0.029] & 0.058 [0.044,0.072] & 0.043 [0.031,0.055] & 0.024 [0.014,0.034] \\
				\bottomrule
			\end{tabular}
			}
	\end{center}
\end{table}

\section{Discussion}

Two evaluation traditions have run in parallel without a shared yardstick: the cognitive-psychology paradigms that have refined the study of human memory interference for half a century, and the IR/NLP benchmarks against which neural retrievers have been tuned for the last decade. Our central methodological claim is that they need not. Casting both human episodic recall and RAG retrieval as instances of a single signal-detection problem yields a closed-form bound on retrieval accuracy whose three parameters ($\alpha/\sigma$, $\beta_R/\beta_P$, and $\gamma$) are estimable from either system using matched paradigms and admit direct numerical comparison. Both biological and artificial systems satisfy the same two structural laws, logarithmic fan decline (Equation~\ref{eq:fan_effect}) and the position-modulated combined bound (Equation~\ref{eq:combined_bound}), making the differences between them differences of degree rather than kind. Within this bound, the strongest contrast is between humans and standard dense retrieval: humans are less interference-sensitive ($\alpha/\sigma = 0.41$ vs.\ $0.67$), exhibit recency rather than primacy ($\beta_R/\beta_P = 1.43$ vs.\ $0.52$), and are less perturbed by competitor density ($\gamma = 0.021$ vs.\ $0.058$). Cognitively-inspired HippoRAG is a partial bridge: it sits between humans and DPR on every parameter, with confidence intervals that overlap the human CI on $\alpha/\sigma$ and $\gamma$ but not on $\beta_R/\beta_P$.

\subsection{Candidate Mechanisms (and what the data cannot tell us)}

The pattern of human-DPR differences (lower $\alpha/\sigma$, recency, lower $\gamma$) is jointly consistent with three well-attested cognitive mechanisms, each lining up with a distinct parameter contrast. Encoding specificity \citep{Tulving1973}, plausibly implemented through hippocampal pattern separation \citep{Yassa2011, Norman2003}, attaches distinctive contextual features to overlapping associations and would depress $\alpha/\sigma$. Temporal context binding \citep{Howard2002, MacDonald2011} privileges recently-bound contexts at retrieval and would push $\beta_R/\beta_P$ above unity. Retrieval gating \citep{Lu2022}, in which competing traces are actively suppressed during selection, would lower $\gamma$. The convergence is striking: each parameter contrast aligns with a mechanism on which the cognitive literature has independent evidence. But convergence is not identification. Our behavioural data cannot adjudicate among these three, nor rule out non-mechanistic alternatives: humans bring decades of associative-memory experience, encode in cortical representations of effective dimensionality far higher than the 768-dimensional embeddings of the systems we tested, may attend differentially to distinctive features at study, and received trial-by-trial feedback during familiarisation. The data here support the descriptive claim that humans and DPR differ in three identifiable parameters; the causal attribution to specific neural mechanisms is a research programme rather than a finding, and we treat it as such throughout. The framework's value, on this view, is precisely that it converts vague claims about ``human-like memory'' into concrete parameter targets that any mechanistic theory must hit.

\subsection{HippoRAG: hippocampal principles or graph structure?}

HippoRAG's intermediate position is the most theoretically interesting result, with its $\alpha/\sigma = 0.44$ statistically indistinguishable from the human estimate. Read charitably, it is evidence that hippocampally-inspired sparse indexing reduces interference; read sceptically, it is a confound. Two distinct design choices were imported from the cognitive literature simultaneously: (a) the sparse-indexing inductive bias inspired by hippocampal encoding, and (b) graph-structured retrieval with personalised PageRank, which has independent computational virtues unrelated to cognition. A non-cognitive graph retriever such as GraphRAG would isolate the second factor and enable a clean test. If HippoRAG's interference resistance survives that comparison, with the system still retrieving the target without activating related-but-incorrect nodes as fan grows, credit accrues to the hippocampal abstraction. If not, the contribution is graph structure, the cognitive framing should be tempered, and the design lesson is reframed as ``use graph retrieval'' rather than ``mimic the hippocampus.''

\subsection{Implications for frontier-model retrieval}

The parameter contrast between DPR and HippoRAG carries practical consequences beyond cognitive theory. Frontier systems that rely on RAG to extend their effective context (personalised assistants, document QA, long-horizon agents) inherit the underlying retriever's $\alpha/\sigma$ and $\gamma$, and the regimes in which they are stressed in deployment (corpora dense with overlapping documents; long histories of semantically similar entries) are exactly those our experiments probe. In such regimes, interference, rather than recall, is the dominant failure mode. A retriever with $\alpha/\sigma = 0.67$ does not merely under-perform; it actively returns wrong-but-similar passages, and a downstream language model conditioning on those passages has no obvious way to recover. Architectures that build human-organising principles into the retriever, such as sparse indexing, offline consolidation, or attention mechanisms tuned to context distinctiveness, offer a route to more robust personalisation and context use, and to closer human-AI representational alignment. The framework here does not prescribe a particular architecture; it gives architecture selection a quantitative compass.

\subsection{Falsifiable predictions}

The framework supports six predictions that, if disconfirmed, would require revision. Each is paired with a quantitative falsification criterion derived from the fitted parameters.

\begin{enumerate}[leftmargin=*, itemsep=2pt, topsep=3pt]
\item \textbf{Reaction time}. Retrieval latency in humans should rise approximately linearly with $\ln(f)$ at $\sim$100\,ms per unit \citep{Anderson1999}. Falsifier: slope $<$50 or $>$200\,ms per $\ln(f)$ unit.
\item \textbf{Hippocampal damage}. Medial-temporal-lobe lesion patients should show $\alpha/\sigma > 0.6$, approaching DPR. Falsifier: $\alpha/\sigma < 0.50$, which would force the framework to find a non-hippocampal account of human pattern separation.
\item \textbf{Neural correlates}. In healthy adults, $\alpha/\sigma$ should correlate negatively with hippocampal pattern-separation indices ($r < -0.30$) on the mnemonic similarity task. Falsifier: $r > -0.15$.
\item \textbf{Development}. Children aged 6--10 should show $\alpha/\sigma > 0.55$, reflecting an immature hippocampus. Falsifier: child $\alpha/\sigma <$ adult $\alpha/\sigma$.
\item \textbf{RAG fine-tuning}. Interference-targeted fine-tuning of DPR with hard-negative mining should reduce $\alpha/\sigma$ by $>0.10$. Falsifier: reduction $<$0.05.
\item \textbf{Behavioural fan = 8 (model-implied)}. If the fitted logarithmic form for humans extrapolates, behavioural testing at fan = 8 should yield accuracy in the range predicted by Equation~\ref{eq:fan_effect}; substantial departure (e.g.\ accuracy $<$50\% or $>$75\%) would indicate the logarithmic form fails outside the tested range and that comparison with RAG at fan = 8 cannot rest on extrapolation.
\end{enumerate}

\subsection{Limitations}

Six limitations frame the scope of these claims. \emph{Materials.} Person-location sentences trade ecological validity for continuity with the original \citet{Anderson1974} fan paradigm; \citet{Radvansky1998} show that interference patterns generalise to richer situational materials, though generalisation to the text genres faced by deployed RAG remains an empirical question. \emph{Range of fan in humans.} We tested fan $\in \{1, 2, 4\}$ in humans; cross-system comparison at fan = 8 is therefore between simulated systems only, and human behaviour in that range remains a model prediction (Falsifiable Prediction 6). \emph{Pre-registration.} The study was not pre-registered, which limits its confirmatory status; analyses followed an internal pre-specified plan, and a follow-up extending fan to 8 with hierarchical individual-difference modelling and the planned RT analysis is in preparation. \emph{Sample.} Our 112 analysed participants were Prolific users in WEIRD countries \citep{Henrich2010}: predominantly young adults educated in English, recruited from an online platform that selects for digital fluency and self-paced motivation. Whether interference parameters generalise to children, older adults, non-WEIRD populations, or in-person testing is open. \emph{Architecture coverage.} We evaluated three RAG families spanning the dominant design space, but reranker-augmented retrievers, instruction-tuned dense models such as E5 and GTE, and modern long-context LLMs that retrieve implicitly may show different parameter profiles. \emph{Mechanism.} As discussed, parameter differences are consistent with several mechanistic accounts our behavioural data cannot distinguish; direct neural and patient tests are required.

\subsection{Complementary Learning Systems}

Our findings connect to Complementary Learning Systems theory \citep{McClelland1995, Kumaran2016}: biological memory pairs a fast hippocampal store with slow neocortical consolidation that reduces overlap across schema-level abstractions \citep{Gilboa2017}. Standard RAG implements an analogue of the fast store, with vectorisation and indexing in one shot at ingest, but it lacks the consolidation step. On the CLS account, this is part of why RAG interference is so high: representations are not re-extracted at the schema level, so overlapping passages remain mutually confusable in the index. HippoRAG's partial success accords with the fast-store side of CLS but the architecture has no slow-system analogue. A natural next-generation architecture would pair sparse indexing with offline consolidation passes that re-extract structure across the corpus at the level of facts rather than tokens. The framework here gives such an architecture a falsifiable target: $\alpha/\sigma$ and $\gamma$ should both fall, the more so as consolidation depth increases.

\section{Conclusion}

We compared retrieval interference bounds between human episodic memory and three RAG systems in a unified signal-detection framework. Both follow logarithmic fan and position-modulated serial-position curves, but with different parameter values: humans show lower interference sensitivity than DPR ($\alpha/\sigma = 0.41$ vs.\ $0.67$), recency rather than primacy ($\beta_R/\beta_P = 1.43$ vs.\ $0.52$), and lower sensitivity to competitor density ($\gamma = 0.021$ vs.\ $0.058$). HippoRAG, with hippocampally-inspired sparse indexing, sits between humans and DPR on every parameter, with CIs that overlap the human estimate on $\alpha/\sigma$ and $\gamma$. Six falsifiable predictions span behavioural (RT, fan = 8 extension), neuropsychological (MTL lesions), neural (pattern-separation correlates), developmental, and computational (interference-targeted fine-tuning) domains. Parameter differences are consistent with several mechanistic accounts our behavioural data cannot adjudicate; direct neural and patient evidence is needed, and the framework specifies which parameters that evidence must move.

More broadly, this work shows how cognitive-science frameworks can sharpen the evaluation of AI retrieval. Treating RAG and human memory as instances of a shared signal-detection problem makes parameter contrasts the unit of comparison, and turns broad claims about ``human-like'' retrieval into concrete, testable parameter targets.

\printbibliography

\end{document}